\documentclass[11pt]{article}
\usepackage{cite}
\usepackage{array}
\usepackage{graphicx}
\usepackage{textpos}
\usepackage[utf8x]{inputenc} 
\usepackage[T1]{fontenc}
\usepackage{graphicx,epstopdf,color} 
\usepackage{amsfonts}
\usepackage{amsmath,amssymb,mathrsfs}
\usepackage{bm}
\usepackage[final,inline]{showlabels} 
\usepackage{mdframed,enumitem}
\usepackage{ifpdf}
\usepackage[normalem]{ulem}
\usepackage{xcolor}

\newcommand{\ii}{\text{i}}
\newcommand{\del}{\partial}
\newcommand{\bdel}{\bar\partial}
\newcommand{\bz}{\bar z}

\newcommand{\bs}[1]{\boldsymbol{#1}}

\newcommand{\x}{\text{x}}
\newcommand{\y}{\text{y}}
\newcommand{\z}{\text{z}}
\newcommand{\of}[1]{\!\left(#1\right)}
\newcommand{\bigof}[1]{\!\bigl(#1\bigr)}
\newcommand{\sqof}[1]{\left[#1\right]}
\newcommand{\cuof}[1]{\left\{#1\right\}}

\newcommand{\bigcomm}[2]{\bigl[#1,#2\bigr]}

\newcommand{\biganticomm}[2]{\bigl\{#1,#2\bigr\}}

\def\ie{{i.e.},\ }

\def\etal{{et al.}}
\def\etc{{etc.}\ }

\title{Adiabatic Construction of Hierarchical Quantum Hall States}
\author{Martin Greiter\\
\small\it Institute for Theoretical Physics, University of  Würzburg, \\
\small\it Am Hubland, 97074 Würzburg, Germany\\[10pt]
Frank Wilczek \\
\small\it Center for Theoretical Physics, MIT, Cambridge, MA 02139 USA; \\
\small\it T. D. Lee Institute and Wilczek Quantum Center, \\
\small\it Shanghai Jiao Tong University, Shanghai, China;\\
\small\it Arizona State University, Tempe, AZ, USA; \\
\small\it Stockholm University, Stockholm, Sweden} 

\begin{document}
\maketitle 

\begin{textblock*}{5cm}(11cm,-10.2cm) 
\fbox{\footnotesize MIT-CTP/5302}
\end{textblock*}

\begin{abstract} 
 We propose an exact model of anyon ground states including higher Landau levels, and use it to obtain fractionally quantized Hall states at filling fractions $\nu=p/(p(m-1)+1)$ with $m$ odd, from integer Hall states at $\nu=p$ through adiabatic localization of magnetic flux.  For appropriately chosen two-body potential interactions, the energy gap remains intact during the process.  The construction hence establishes the existence of incompressible states at these fillings.  
\end{abstract}

%
%

\emph{Introduction}.---The adiabatic principle for quantized Hall states\cite{greiter-90mplb1063} asserts that such states at different Landau filling fractions\cite{laughlin83prl1395,haldane83prl605,halperin84prl1583,arovas-84prl722,jain89prl199,greiter94plb48,Wen04,hansson-07prb075347,bonderson12prl066806,hansson-17rmp025005,tournois-20spp79} can be obtained from each other through a process of adiabatic localization of magnetic flux onto the particles\cite{wilczek82prl1144,wilczek82prl957,girvin-87prl1252,read89prl86,Wilczek90}.  Using an exactly solvable model of anyons\cite{girvin-90prl1671,jackiw-90prl2969,johnson-90prb6870}, we have previously\cite{greiter-92npb577} been able to interpolate continuously between a filled Landau level and Laughlin $1/m$ states\cite{laughlin83prl1395}, as well as a p-wave superfluid and a Pfaffian state at Landau level filling fraction $\nu=\frac{1}{2}$\cite{moore-91npb362,greiter-91prl3205,greiter-92npb567,nayak-96npb529,read-00prb10267,ivanov01prl268,levin-07prl236806,lee-07prl236807,son15prx031027,rezayi17prl026801,simon-20prb041302,sun-20prb121303}.  In particular, we were able to show that during the process of trading uniform flux for flux localized on the particles, the energy gap in the spectrum remains intact.  Very recently Hansson and Kivelson \cite{hansson-preprint} have proposed a different implementation of the same strategy, based on interpolating between strong and weak coupling in an auxiliary gauge field, which has an elegant field-theoretic formulation.  

In view of fruitful work of Jain and collaborators\cite{jain89prl199,
  jain-90prl1297,Jain07}, offering good variational wave functions for a wider variety of fractional quantum Hall states, based on constructions involving several Landau levels, it is natural to ask whether our analytical model allows generalizations that take off from several filled Landau levels.  If the initial number of filled Landau levels is $p$, then localization of $\,2\theta/\pi=(m-1)$ Dirac flux quanta on each particle results in a state with inverse filling fraction $1/\nu=1/p + (m-1)$.  Here $m$ must be an odd integer for the quantum statistics of the final particles to evolve back to the statistics of the initial particles.  The simplest fractions, $2/5,\, 3/7$, and $4/9$, are often referred to as the Jain series of hierarchy states. In this work, we obtain them by continuously deforming one incompressible quantum Hall state into another.  At intermediate stages we have states of anyons in a background magnetic field.  If a finite gap remains intact during the process---as it does for specific families of local Hamiltons---then the construction establishes the existence of incompressible many-body fermion states at the final fractions, and provides explicit wave functions.

In this note, we first review our earlier work, while also extending it slightly to include isolated, fractional flux tubes not localized on any of the particles.  Secondly, we generalize the exact model to higher Landau levels, where it will yield an analytic derivation of quantized Hall states at the Jain fractions.  During the adiabatic localization of flux, the energy gap is assumed to remain intact, as we will explicitly show for the $p=2$ series.  The wave functions we obtain through this process are equivalent to the {\em un}projected wave functions proposed (in projected form) by Jain.  Our theory establishes the existence of a universality class of incompressible states at those fractions.  It is plausible, in view of experimental observations and the numerical work mentioned above, that this universality class includes realistic physical systems.  Similarly to what has been done in QCD, where one connects the solvable (gapped, confining) ``strong coupling'' limit to the real world, one might hope to demonstrate this numerically, by deforming the model parent Hamiltonians at the end of our adiabatic process to realistic ones along a path that maintains a finite gap.

\emph{Review and extension of the lowest Landau level model}.---Our earlier work was broadly based on a model of anyons in the lowest Landau level (LLL) proposed by Girvin \etal\cite{girvin-90prl1671} and Jackiw and Pi\cite{jackiw-90prl2969}. Here we briefly review this model, and extend it so as to include a fractional flux tube not attached to any of the particles.  The generalization to an arbitrary number of such flux tubes involves no new ideas.  We use units where $\hbar=c=1$.

To begin, consider the following Hamiltonian of $N$ particles in the plane spanned by unit vectors $\bs{\hat e}_\x$ and $\bs{\hat e}_\y$,
\begin{align}
  \label{eq:Hphys} 
  H=\frac{1}{2m}~\sum_{i=1}^N \sqof{\of{-\ii\nabla_i+e\bs{A}_i}^2-eB_i},
\end{align}
where particle $i$ is coupled to a magnetic field perpendicular to the plane,
\begin{align}
  \label{eq:Biphys} 
  \bs{B}_i=\nabla_i\times\bs{A}_i=:-B_i\bs{\hat e}_\z.
\end{align}
Introducing complex coordinate notation $z\equiv x+\ii y$, $\del=\frac{1}{2}\of{\del_\x-\ii\del_\y}$, $A=A_\x+\ii A_\y$, and the ladder-like operators
\begin{align}
  \label{eq:aa} 
  a_i^{\phantom{\dagger}}\equiv \sqrt{2}\of{\bdel_i + \frac{\ii e}{2}A_i}, \quad
  a_i^{\dagger}=\sqrt{2}\of{-\del_i - \frac{\ii e}{2}\bar A_i}, 
\end{align}
we find the (anti-)commutations relations
\begin{align}
  \biganticomm{a_i^{\phantom{\dagger}}}{a_i^\dagger}
  =\left(-\ii\nabla_i+e\bs{A}_i\right)^2,\quad
  \bigcomm{a_i^{\phantom{\dagger}}}{a_i^\dagger}
  =\ii e \of{\del_i A_i -\bdel_i \bar A_i} 
  =eB_i,
\end{align}
and hence 
\begin{align}
  \label{eq:Haa} 
  H=\frac{1}{m}~\sum_{i=1}^N a_i^\dagger a_i^{\phantom{\dagger}}.
\end{align}
Since $H$ is positive semi-definite, every state annihilated by all the operators $a_i$ is a zero energy ground state.  For a uniform magnetic field $B_i=B$, the above formalism trivially reproduces Landau level quantization.

For the present analysis, we take 
\begin{align}
  \label{eq:Bi}
  B_i=-\frac{\varphi}{e}\delta^2(\bs{r}_i-\bs{\eta})
  -\frac{2\theta}{e}\sum_{j(\not= i)}\,\delta^2(\bs{r}_i-\bs{r_j})\,+\,B,
\end{align}
where $B$ is a uniform magnetic field pointing in the negative 
$\bs{\hat e}_\z$-direction, while the first two terms describe (fractional) flux tubes at positions $\bs{\eta}$ and $\bs{r}_j$ pointing in the positive $\hat\z$-direction.  We choose the symmetric gauge $\nabla_i\bs{A}_i=\del_i A_i +\bdel_i \bar A_i=0$.  Then \eqref{eq:Bi} is generated by
\begin{align}
  \label{eq:Ai}
  \bs{A}_i
  =-\frac{\varphi}{2\pi e}
  \frac{(\bs{r}_i-\bs{\eta})
     \times\bs{e}_\z}{\left|\bs{r}_i-\bs{\eta}\right|^2}
  -\frac{\theta}{\pi e}\,\sum_{j (\not= i)}\,
  \frac{(\bs{r}_i-\bs{r}_j)\times\bs{e}_\z}{\left|\bs{r}_i-\bs{r}_j\right|^2}
  +\frac{1}{2}B(\bs{r}_i\times \bs{e}_\z).
\end{align}
In complex coordinate notation, we write
\begin{align}
  \label{eq:AiBicomplex}
  A_i=-\frac{2\ii}{e}\bdel_i S,\quad \bar A_i=\frac{2\ii}{e}\del_i S,\quad 
  B_i=\frac{4}{e}\del_i\bdel_iS,
\end{align}
where we have introduced
\begin{align}
  \label{eq:S}
  S~\equiv~\frac{\varphi}{2\pi}\sum_i\ln|z_i-\eta| 
  +\frac{\theta}{\pi}\sum_{i<j}\ln|z_i-z_j| 
  + \frac{1}{4}eB\sum_{i}|z_i|^2.
\end{align}
The equivalence of \eqref{eq:AiBicomplex} with \eqref{eq:S} to \eqref{eq:Bi} and \eqref{eq:Ai} is easily verified with the identity
\begin{align}
  \label{eq:2}
  \del\bdel\ln(z\bz)=
  \del\,\frac{1}{\bz}=\bdel\,\frac{1}{z}=\pi\delta^2(z),
\end{align}
where the $\delta$-function is understood to be taken over real and imaginary part separately.
Using $S$, it is convenient to rewrite the ladder-like operators \eqref{eq:aa} as
\begin{align}
  \label{eq:aS}
  a_i^{\phantom{\dagger}}&=+\sqrt{2}\, e^{-S}\,\bdel_i\, e^{+S}, \quad\\[5pt]
  \label{eq:aSdag}
  a_i^{\dagger}&=-\sqrt{2}\, e^{+S}\,\del_i\, e^{-S}
  =\sqrt{2}\, e^{-S}\bigof{-\del_i+2(\del_iS)} e^{+S}. 
\end{align}
This clearly shows that the manifold of exact zero energy ground states of
\eqref{eq:Hphys} or \eqref{eq:Haa} with \eqref{eq:Bi} and \eqref{eq:Ai} is spanned by
\begin{align}
  \label{eq:psi0}
  \psi_0[z]=f[z]\,e^{-S},
\end{align}
where $f[z]$ is an arbitrary entire function of the complex particle
positions.  These states are, of course, not always normalizable.

The model describes charged particles with magnetic flux tubes attached in a uniform magnetic field $B$.  The flux attached to the particles is given by $2\theta /e$, and the particles see an additional flux tube 
of strength 
$\varphi /e$ at complex position $\eta =\eta_\x+\ii\eta_y$.  With regard to the uniform magnetic field, the ground states \eqref{eq:psi0} are in the lowest Landau level.

The model is relevant for the adiabatic principle if we choose a filled Landau level as initial state for $\theta=\varphi=0$, \ie 
\begin{align}
  \label{eq:fLLL}
  f[z] =\prod_{i<j}\,(z_i-z_j).
\end{align}
During the process of adiabatic localization of flux, the zero energy ground state is then given by
\begin{align}
  \label{eq:psi}
  \psi^{(\varphi,\theta)}[z] =\prod_{i<j}\,(z_i-z_j)\,
  \prod_{i}\,|z_i-\eta|^{\varphi/2\pi}\,
  \prod_{i<j}\,|z_i-z_j|^{\theta/\pi}\,
  \prod_i e^{-\frac{1}{4}eB |z_i|^2}.
\end{align}
For $\varphi=0$, this state gradually evolves from a filled Landau level to a state reminiscent of a Laughlin $1/m$ state as we evolve the statistics from fermions to superfermions by taking $\theta$ from $0$ to $(m-1)\pi$.  To fully recover the Laughlin states, we need to remove the flux attached to the particles via a ``singular gauge transformation'',
\begin{align}
  \label{eq:singau}
  \bs{A}_i \to 
  \bs{A}_i+\nabla_i\Lambda_i, \quad 
  \psi[z] \to 
  \psi[z] \,\prod_i e^{-\ii e\Lambda_i},
\end{align}
with
\begin{align}
  \label{eq:lam}
  \Lambda_i=-\frac{\theta}{2\pi e}\,\sum_{j (\ne i)}\hbox{arg}(z_i-z_j).
\end{align}
It is called ``singular'' because it is ill-defined when the particle positions coincide.  For $\theta\ne 0\!\mod 2\pi$, however, this possibility is excluded due to the angular momentum barrier between the particles.  For $\theta=(m-1)\pi$, where $m=3,5,7,\ldots$ is an odd integer, \eqref{eq:psi} tranforms into
\begin{align}
  \label{eq:finalpsi}
  \psi^{(\varphi,m)}[z] =
  \prod_{i}\,|z_i-\eta|^{\varphi/2\pi}\,
  \prod_{i<j}\,(z_i-z_j)^m\,
  \prod_i e^{-\frac{1}{4}eB |z_i|^2}.
\end{align}
When $\varphi$ is a multiple of $2\pi$, we may likewise remove the isolated flux tube at $\eta$ via another similar gauge transformation. As discussed in previous work\cite{greiter-92npb577}, the energy gap remains intact during this process if we supplement the parent Hamiltonian \eqref{eq:Hphys} with a repulsive interaction of the form
\begin{align}
  \label{eq:V_theta}
  V^{(1+\theta/\pi)}
  \,\propto\,\sum_{i\ne j}\, |z_i - z_j|^{2n-\theta/\pi}\,
  \left(\nabla_i^2 \right)^n\,\delta^2(z_i - z_j)\, .
\end{align}
where n is an integer such that $2n-\theta/\pi\ge 0$.

There is a subtlety associated with the singular gauge transformation.  If we view it as only a gauge transformation which is ill defined at a set of points we may safely exclude form our configuration space (the points where at least two particle positions coincide), it will not affect the $\delta$-function interaction in \eqref{eq:Hphys}.  This can be seen from application of \eqref{eq:singau} to \eqref{eq:Hphys} and \eqref{eq:Biphys}.  The gauge transformation removes the effect of the flux tubes in the kinetic term of \eqref{eq:Hphys}, but does not affect the potential term $-eB_i$.
Since a prerequisite for the singular gauge transformation is that we exclude configurations where particle positions coincide, however, we are permitted to omit the $\delta$-function interactions from the Hamiltonian once all the flux tubes are gauged away.  When we say that we remove the flux tubes when $\theta$ is a multiple of $\pi$ via a singular gauge transformation, we from now on take this to mean that we transform the wave function according to \eqref{eq:singau}, and (in the absence of additional flux tubes parametrized by $\varphi$) replace 
$a_{i}^{\phantom{\dagger}}, a_{i}^{\dagger}$ by the (unnormalized) Landau level operators 
\begin{align}
  \label{eq:a0S0}
  a_{0i}^{\phantom{\dagger}}=+\sqrt{2}\, e^{-S_0}\,\bdel_i\, e^{+S_0},\quad 
  a_{0i}^{\dagger}=-\sqrt{2}\, e^{+S_0}\,\del_i\, e^{-S_0},
\end{align}
with $S_0~=~\frac{1}{4}eB\sum_{i}|z_i|^2$
in \eqref{eq:Haa} to obtain
\begin{align}
  \label{eq:H0aa} 
  H_0=\frac{1}{m}~\sum_{i=1}^N a_{0i}^\dagger a_{0i}^{\phantom{\dagger}}.
\end{align}

\emph{Generalization of the model to higher Landau levels}.---We proceed by generalizing the model to an initial state consisting of $p$ filled Landau levels, where $p$ is a positive integer.  Since the wave function for a filled $q$-th Landau level is obtained from the wave function of a filled LLL by application of $(q-1)$ Landau level raising operators $a_{0i}^{\dagger}$ for each particle, the wave function for $p$ filled levels 
%
may be written as
\begin{align}
 \label{eq:psi0_p}\nonumber
 \psi_p^{(0)}[z]
 &={\mathcal A}\,\prod_{q=1}^p\cuof{\prod_{i=(q-1)N_1+1}^{qN_1} (\bz_i)^{q-1}
 \prod_{(q-1)N_1<i<j\le qN_1}(z_i-z_j)} e^{-S_0}\\[10pt]
 &=: f_p[z,\bz]\, e^{-S_0},
\end{align}
where ${\mathcal A}$ denotes antisymmetrization, the total number of particles $N$ is divisible by $p$, $N_1=N/p$ is the number of particles in each Landau level, and $f_p[z,\bz]$ abbreviates the polynomial part of the preceding line.
$\psi_p^{(0)}[z]$ is clearly the ground state of 
\eqref{eq:H0aa} 
%
at the appropriate Landau level filling $\nu=p$, but, even if we replace the factor $e^{-S_0}$ by $e^{-S}$, not an eigenstate of \eqref{eq:Haa}
for $\theta\not= 0$.  (From here onwards, we simplify the presentation by taking $\varphi =0$, even though additional flux tubes may be included without incident.)

Therefore, the question arises whether we can formulate a parent Hamiltonian for 
\begin{align}
  \label{eq:psi_p} 
  \psi_p^{(\theta)}[z]=: f_p[z,\bz]\, e^{-S},
\end{align}
with $S$ given by \eqref{eq:S} with $\varphi =0$, which singles out $\psi_p^{(0)}[z]$ as unique ground state for $\theta >0$ as well.

This is indeed possible, and simpler than it may appear.  The Hamiltonian in question is given by
\begin{align}
  \label{eq:Hpaa} 
  H_p^{(\theta)} = \frac{1}{m(eB)^{p-1}} \sum_{i=1}^N
  \bigl(a_{i}^\dagger\bigr)^p\bigl(a_{i}^{\phantom{\dagger}}\bigr)^p.
\end{align}
The factor $(eB)^{p-1}$ in the denominator ensures that $H_p^{(\theta)}$ has the correct units.  To verify that \eqref{eq:psi_p} is a zero energy ground state, we note that $H_p^{(\theta)}$ is positive semi-definite, and use \eqref{eq:aS} and \eqref{eq:aSdag} to rewrite it as \begin{align}
  \label{eq:Hpdd} 
  H_p^{(\theta)} = \frac{2^p}{m(eB)^{p-1}} \sum_{i=1}^N  
  e^{-S}\,\bigl(-\del_i+2(\del_iS)\bigr)^p\,\bigl(\bdel_i\bigr)^p\, e^{+S}.
\end{align}
Since $f_p[z,\bz]$ is the unique antisymmetric polynomial in $N$ complex coordinates $z_i,\bz_i$ containing powers up to $N_1-1$ in the $z_i$'s and powers up to $p-1$ in the $\bz_i$'s, the state \eqref{eq:psi_p} constitutes the unique zero energy ground state of \eqref{eq:Hpdd} at Landau level filling fraction $\nu=p$.  All states with weight in higher Landau levels contain powers $\bz_i^p$ or higher, and are hence not annihilated by the derivatives $\bdel_i$ in \eqref{eq:Hpdd}.

The leading kinetic term in \eqref{eq:Hpaa} is $(-\ii\nabla_i)^{2p}$, which is hard to realize in a material system, but not unphysical as a matter of principle. The Hamiltonian also contains products of derivatives and $\delta$-functions, but these terms matter only when particle positions coincide, \ie at points we will be allowed to exclude from the configuration space due to the angular momentum barrier during the adiabatic process.  The important feature in the present context is that all the derivative terms are minimally coupled to the gauge field $\bs{A}_i$, as one can easily see from \eqref{eq:aa}, where the derivatives enter exclusively in combinations $(-\ii\del_\x+eA_\x)$ and $(-\ii\del_\y+eA_\y)$.  Therefore, the model describes flux tube--particle composites subject to a uniform magnetic field $B$, with an unusual kinetic energy for the particles.

The Hamiltonian \eqref{eq:Hpaa} simplifies when $\theta =0$, \ie at the point of departure for the adiabatic process, and also after the flux tubes have been removed via the singular gauge transformation.  Then $a_{i}^{\phantom{\dagger}}, a_{i}^{\dagger}$ may be replaced by $a_{0i}^{\phantom{\dagger}}, a_{0i}^{\dagger}$, and 
\eqref{eq:Hpaa} may be written as
\begin{align}
 \label{eq:H0paa} \nonumber 
 H_p^{(0)} &= \frac{1}{m(eB)^{p-1}}\sum_{i=1}^N\; \prod_{q=1}^p 
 \of{a_{0i}^\dagger a_{0i}^{\phantom{\dagger}} -(q-1)eB} \\
       &= \frac{1}{2^pm(eB)^{p-1}}\sum_{i=1}^N\; \prod_{q=1}^p 
 \Bigl(\of{-\ii\nabla_i+e\bs{A}_i}^2 -(2q-1)eB\Bigr). 
\end{align}
which shows that it may be viewed as a sum over products of projection operators, and that it contains the factor $\of{-\ii\nabla_i+e\bs{A}_i}$ taken to all even powers up to $2p$. 
%
%

The ground state \eqref{eq:psi_p} of \eqref{eq:Hpaa} during the adiabatic localization of flux parametrized by $\theta$ is given by
\begin{align}
  \label{eq:psi_p_noS}
  \psi_p^{(\theta)}[z] =  \prod_{i}\,|z_i-z_j|^{\theta/\pi}\,
   f_p[z,\bz]\,\prod_i e^{-\frac{1}{4}eB |z_i|^2}.
\end{align}
It gradually evolves from a state with $p$ filled Landau levels into states reminiscent of unprojected Jain states at filling fractions 
$\nu=p/(p\,(m-1)+1)$ as we evolve the statistics from fermions to superfermions by taking $\theta$ from $0$ to $(m-1)\pi$, where $m=3,5,7,\ldots$ is an odd integer.  To fully recover the unprojected Jain states, 
\begin{align}
  \label{eq:unprojected_Jain}
  \psi_p^{(m)} [z] =  \prod_{i}\,(z_i-z_j)^{m-1}\,
   f_p[z,\bz]\,\prod_i e^{-\frac{1}{4}eB |z_i|^2},
\end{align}
we need to remove the flux attached to the particles with the singular gauge transformation \eqref{eq:singau} with \eqref{eq:lam}.  Following the argument outlined above, we are allowed to discard the flux entirely whenever $\theta$ is a multiple of $\pi$.  The 
Hamiltonian hence evolves back into its initial form \eqref{eq:H0paa}.

While it is obvious that \eqref{eq:psi_p_noS} is an exact zero energy ground state of \eqref{eq:Hpaa} during the process, we only know it to be non-degenerate at the initial point $\theta=0$, where it is separated from all other states by a gap in the kinetic energy.

During the process, states containing inverse powers $(z_i-z_j)^{-1}$ of particle distances become normalizable, as elaborated in reference \cite{greiter-92npb577}, and we expect that \eqref{eq:psi_p_noS} is no longer the only ground state.  Fortunately, the line of arguments developed there can be applied to the present model as well.  First of all, note that $f_p[z,\bz]$ vanishes at least linearly as two particles approach each other.  This follows from the polynomial form together with the antisymetrization, which is required by Fermi statistics.  Taking this argument one step further, the unprojected Jain state \eqref{eq:unprojected_Jain} vanishes by construction as the $m$-th power of the distance as two particles approach each other.  The potential term introduced by Trugman and Kivelson\cite{trugman-85prb5280} for the $\nu=1/m$ Laughlin state in the LLL, \begin{align}
  \label{eq:V_m}
  V^{(m)}
  \,\propto\,\sum_{i\ne j}\,\of{\nabla_i^2}^{(m-1)/2}\delta^2(z_i-z_j),
\end{align}
hence also annihilates the unprojected Jain state \eqref{eq:unprojected_Jain}.
If we combine \eqref{eq:V_m} with the kinetic Hamiltonian \eqref{eq:H0paa},
\begin{align}
  \label{eq:H_unprojected_Jain}
  H_p^{(m)} = H_p^{(0)}+V^{(m)}, 
\end{align}
we thus obtain a Hamiltonian\cite{rezayi-91prb8395,chen-17prb195169} which annihilates the unprojected Jain state \eqref{eq:unprojected_Jain} at the appropriate filling fraction.  Unfortunately, \eqref{eq:unprojected_Jain} is singled out as the unique ground state only for $p=2$, corresponding to filling fractions $\nu=2/5$, $2/9$, \etc  The reason is that for $p\ge 3$, parton states\cite{jain89prb8079,bandyopadhyay-18prb161118,faugno-19prl016802,kim-19np154
} with higher electron densities are also annihilated by \eqref{eq:H_unprojected_Jain}.  A construction of parent Hamiltonians for the full series of unprojected Jain states has been accomplished recently by Bandyopadhyay \etal\cite{bandyopadhyay-20prl196803}.  It appears that this family of Hamiltonians can be extended along the lines we used to derive \eqref{eq:V_theta} from \eqref{eq:V_m}, and thus also provides the required potentials for the energy gap to remain intact during the adiabatic process.

These considerations imply that if we supplement \eqref{eq:Hpaa} with the potential term \eqref{eq:V_theta} for $p=2$, and suitable generalizations for higher fractions, the state $\psi_p^{(\theta)}[z]$ given by \eqref{eq:psi_p_noS} will remain the exact and non-degenerate ground state during the process.  The kinetic gap of the initial state continuously evolves into a gap due to repulsive interactions, which are characteristic in stabilizing fractionally quantized Hall states.  The principle establishes why we expect particularly stable Hall states at certain fillings, and provides a strong physical argument for their incompressibility.  This is our main result.

To our knowledge, no parent Hamiltonian has been identified for any projected hierarchical FQHE state.  For states in the LLL, it has even been shown numerically that no two-body parent Hamiltonian exists for standard composite fermion or hierarchy wave functions in finite systems\cite{greiter94plb48}.  Jain, Rezayi, and their respective coworkers\cite{rezayi-91prb8395,jain-97ijmpb,Jain07} have established through numerical work that the states \eqref{eq:unprojected_Jain}, when projected onto the LLL, are excellent trial wave functions for quantized Hall states at filling fractions $2/5$ and $3/7$ with realistic interactions, in the sense of having large numerical overlaps.   Numerical success of specially constructed trial wave functions, however, does not establish the existence of a universality class. 

Projection onto the LLL corresponds to the limit of infinite cyclotron frequency while keeping the magnetic length fixed, and is implemented by taking the mass $m\to 0$.  This is an interesting limit, though it is not physically mandatory.  The form of our kinetic Hamiltonian \eqref{eq:H0paa} seems to suggest such a projection: Since it is a sum over products of operators projecting out individual Landau levels, it is evident that only the projector annihilating states in the LLL is required if we move all other projection operators from the Hamiltonian onto the state.  The problem here is that the projected state is no longer annihilated by the potential term $V^{(m)}$, regardless of whether $V^{(m)}$ is projected into the LLL as well.  For this reason, our analysis does not suggest an avenue towards the identification of parent Hamiltonians for the {\it projected\/} state.
  
In conclusion, let us add two general remarks.  The first is that the easy possibility to include flux tubes as separate degrees of freedom in our procedure opens up possibilities for additional adiabatic constructions of hierarchical ground and quasiparticle states.  The second is that the intermediate states in our construction, in between those that can be interpreted as fermion many-body states, represent incompressible ``hierarchical'' many-body states of anyons in a background magnetic field.


\emph{Acknowledgements:}  We wish to thank Ajit Balram for alerting us to several recent references.  MG is supported by the Deutsche 
Forschungsgemeinschaft (DFG, German Research Foundation)---Project-ID 258499086---SFB 1170, through the Würzburg-Dresden Cluster of Excellence on Complexity and Topology in Quantum Matter---\textit{ct.qmat} Project-ID 390858490---EXC 2147, and through DFG grant GR-1715/3-1.  FW is supported by the U.S. Department of Energy under grant Contract  Number DE-SC0012567, by the European 
Research Council under grant 742104, and by the Swedish Research Council under Contract No. 335-2014-7424.

\renewcommand{\refname}{\normalsize References}


\begin{thebibliography}{10}

\bibitem{greiter-90mplb1063}
M. Greiter and F. Wilczek, Mod. Phys. Lett. B {\bf 4},  1063  (1990).

\bibitem{laughlin83prl1395}
R.~B. Laughlin, Phys. Rev. Lett. {\bf 50},  1395  (1983).

\bibitem{haldane83prl605}
F.~D.~M. Haldane, Phys. Rev. Lett. {\bf 51},  605  (1983).

\bibitem{halperin84prl1583}
B.~I. Halperin, Phys. Rev. Lett. {\bf 52},  1583  (1984), \emph{ibid.}
  \textbf{52}, 2390(E) (1984).

\bibitem{arovas-84prl722}
D. Arovas, J.~R. Schrieffer, and F. Wilczek, Phys. Rev. Lett. {\bf 53},  722
  (1984).

\bibitem{jain89prl199}
J.~K. Jain, Phys. Rev. Lett. {\bf 63},  199  (1989).

\bibitem{greiter94plb48}
M. Greiter, Phys. Lett. B {\bf 336},  48  (1994).

\bibitem{Wen04}
X. Wen, {\em Quantum Field Theory of Many-Body Systems} (Oxford University, New
  York, 2004).

\bibitem{hansson-07prb075347}
T.~H. Hansson, C.-C. Chang, J.~K. Jain, and S. Viefers, Phys. Rev. B {\bf 76},
  075347  (2007).

\bibitem{bonderson12prl066806}
P. Bonderson, Phys. Rev. Lett. {\bf 108},  066806  (2012).

\bibitem{hansson-17rmp025005}
T.~H. Hansson, M. Hermanns, S.~H. Simon, and S.~F. Viefers, Rev. Mod. Phys.
  {\bf 89},  025005  (2017).

\bibitem{tournois-20spp79}
Y. Tournois, M. Hermanns, and T.~H. Hansson, SciPost Phys. {\bf 8},  79
  (2020).

\bibitem{wilczek82prl1144}
F. Wilczek, Phys. Rev. Lett. {\bf 48},  1144  (1982).

\bibitem{wilczek82prl957}
F. Wilczek, Phys. Rev. Lett. {\bf 49},  957  (1982).

\bibitem{girvin-87prl1252}
S.~M. Girvin and A.~H. MacDonald, Phys. Rev. Lett. {\bf 58},  1252  (1987).

\bibitem{read89prl86}
N. Read, Phys. Rev. Lett. {\bf 62},  86  (1989).

\bibitem{Wilczek90}
F. Wilczek, {\em Fractional statistics and anyon superconductivity} (World
  Scientific, Singapore, 1990).

\bibitem{girvin-90prl1671}
S.~M. Girvin, A.~H. MacDonald, M.~P.~A. Fisher, S.-J. Rey, and J.~P. Sethna,
  Phys. Rev. Lett. {\bf 65},  1671  (1990).

\bibitem{jackiw-90prl2969}
R. Jackiw and S.-Y. Pi, Phys. Rev. Lett. {\bf 64},  2969  (1990).

\bibitem{johnson-90prb6870}
M.~D. Johnson and G.~S. Canright, Phys. Rev. B {\bf 41},  6870  (1990).

\bibitem{greiter-92npb577}
M. Greiter and F. Wilczek, Nucl. Phys. B {\bf 370},  577  (1992).

\bibitem{moore-91npb362}
G. Moore and N. Read, Nucl. Phys. B {\bf 360},  362  (1991).

\bibitem{greiter-91prl3205}
M. Greiter, X.~G. Wen, and F. Wilczek, Phys. Rev. Lett. {\bf 66},  3205
  (1991).

\bibitem{greiter-92npb567}
M. Greiter, X.~G. Wen, and F. Wilczek, Nucl. Phys. B {\bf 374},  567  (1992).

\bibitem{nayak-96npb529}
C. Nayak and F. Wilczek, Nucl. Phys. B {\bf 479},  529  (1996).

\bibitem{read-00prb10267}
N. Read and D. Green, Phys. Rev. B {\bf 61},  10267  (2000).

\bibitem{ivanov01prl268}
D.~A. Ivanov, Phys. Rev. Lett. {\bf 86},  268  (2001).

\bibitem{levin-07prl236806}
M. Levin, B.~I. Halperin, and B. Rosenow, Phys. Rev. Lett. {\bf 99},  236806
  (2007).

\bibitem{lee-07prl236807}
S.-S. Lee, S. Ryu, C. Nayak, and M.~P.~A. Fisher, Phys. Rev. Lett. {\bf 99},
  236807  (2007).

\bibitem{son15prx031027}
D.~T. Son, Phys. Rev. X {\bf 5},  031027  (2015).

\bibitem{rezayi17prl026801}
E.~H. Rezayi, Phys. Rev. Lett. {\bf 119},  026801  (2017).

\bibitem{simon-20prb041302}
S.~H. Simon, M. Ippoliti, M.~P. Zaletel, and E.~H. Rezayi, Phys. Rev. B {\bf
  101},  041302(R)  (2020).

\bibitem{sun-20prb121303}
C. Sun, K.~K.~W. Ma, and D.~E. Feldman, Phys. Rev. B {\bf 102},  121303(R)
  (2020).

\bibitem{hansson-preprint}
T. Hansson and S. Kivelson, \emph{{M}ean Field Theories of Quantum Hall Liquids
  Justified: Variations on the Greiter Wilczek Theme}, private communication,
  to appear.

\bibitem{jain-90prl1297}
J.~K. Jain, S.~A. Kivelson, and N. Trivedi, Phys. Rev. Lett. {\bf 64},  1297
  (1990).

\bibitem{Jain07}
J.~K. Jain, {\em Composite Fermions} (Cambridge University Press, Cambridge,
  2007).

\bibitem{trugman-85prb5280}
S.~A. Trugman and S. Kivelson, Phys. Rev. B {\bf 31},  5280  (1985).

\bibitem{rezayi-91prb8395}
E.~H. Rezayi and A.~H. MacDonald, Phys. Rev. B {\bf 44},  8395(R)  (1991).

\bibitem{chen-17prb195169}
L. Chen, S. Bandyopadhyay, and A. Seidel, Phys. Rev. B {\bf 95},  195169
  (2017).

\bibitem{jain89prb8079}
J.~K. Jain, Phys. Rev. B {\bf 40},  8079(R)  (1989).

\bibitem{bandyopadhyay-18prb161118}
S. Bandyopadhyay, L. Chen, M.~T. Ahari, G. Ortiz, Z. Nussinov, and A. Seidel,
  Phys. Rev. B {\bf 98},  161118  (2018).

\bibitem{faugno-19prl016802}
W.~N. Faugno, A.~C. Balram, M. Barkeshli, and J.~K. Jain, Phys. Rev. Lett. {\bf
  123},  016802  (2019).

\bibitem{kim-19np154}
Y. Kim, A.~C. Balram, T. Taniguchi, K. Watanabe, J.~K. Jain, and J.~H. Smet,
  Nat. Phys. {\bf 15},  154–158  (2019).

\bibitem{bandyopadhyay-20prl196803}
S. Bandyopadhyay, G. Ortiz, Z. Nussinov, and A. Seidel, Phys. Rev. Lett. {\bf
  124},  196803  (2020).

\bibitem{jain-97ijmpb}
J.~K. Jain and R.~K. Kamilla, Int. J. Mod. Phys. B {\bf 11},  2621  (1997).

\end{thebibliography}
\end{document}